\documentclass[a4paper,10pt]{article} 
\usepackage[left=.65in,top=.65in,bottom=.65in,right=.65in]{geometry}
\usepackage{setspace} % use this for double-spacing
\usepackage[inline,shortlabels]{enumitem} % in-line enumerated items
\usepackage[super]{cite}

\usepackage{ragged2e}
\usepackage{amsmath}
\usepackage{amsthm}
\usepackage{amssymb}
\usepackage{caption}
\usepackage{multicol}
\usepackage[hidelinks]{hyperref}
\usepackage{nameref}
\usepackage{url}
\usepackage{lmodern}
\usepackage{graphicx} 
\usepackage{booktabs} 
\usepackage[T1]{fontenc}
\usepackage{helvet}         % Loads Helvetica font
\usepackage{setspace}

\usepackage[utf8]{inputenc} % if not already present
\usepackage{newunicodechar}
\newunicodechar{−}{-} % maps Unicode minus to a normal hyphen
\newunicodechar{́}{\'{}}

\newcommand{\muhz}{$\mu$Hz}
\newcommand{\numax}{$\nu_{\mathrm{max}}$}

\newcommand{\dnu}{$\Delta\nu$}

\newcommand{\smalltwo}{$\delta\nu_{0,2}$}

\newcommand{\msol}{M$_\odot$}

\newcommand{\kepler}{\textit{Kepler}}

\newcommand{\apj}{ApJ}
\newcommand{\aap}{A\&A}
\newcommand{\apjs}{ApJS}
\newcommand{\mnras}{MNRAS}
\newcommand{\apjl}{ApJL}
\newcommand{\aj}{AJ}
\newcommand{\aapr}{A\&A~Rev.}
\newcommand{\apss}{Ap\&SS}

\makeatletter
\renewcommand{\maketitle}{\bgroup\setlength{\parindent}{0pt}

    \begin{Large} \begin{spacing}{1} \@title\end{spacing}\end{Large}
    \bigskip\linespread{1.2}
    \begin{normalsize} \@author \end{normalsize}

\bigskip
\@date
}
\makeatother

\newcommand{\beginsupplement}{%
    \setcounter{figure}{0}
}
\usepackage[symbol]{footmisc}

\title{\textbf{Acoustic modes in M67 cluster stars trace deepening convective envelopes}}
\author{%
Claudia Reyes{\textsuperscript{1,2}}, %
Dennis Stello$^{1,3}$, 
Joel Ong$^{4}$\textsuperscript{\textdagger},
Christopher Lindsay${^5}$,
Marc Hon$^{6,4}$, and
Timothy R. Bedding$^{3}$. \\}

\date{\footnotesize  
$^{1}$ School of Physics, University of New South Wales, NSW 2052, Australia.\\
$^{2}$ Research School of Astronomy \& Astrophysics, Australian National University, Canberra ACT 2611, Australia.\\
$^{3}$ Sydney Institute for Astronomy (SIfA), School of Physics, University of Sydney, NSW 2006, Australia.\\
$^{4}$ Institute for Astronomy, University of Hawaii, 2680 Woodlawn Drive, Honolulu, HI 96822, USA.\\
$^{5}$ Department of Astronomy, Yale University, PO Box 208101, New Haven, CT 06520-8101, USA.\\
$^{6}$ Department of Physics and Kavli Institute for Astrophysics and Space Research, Massachusetts Institute of Technology, 77 Massachusetts Ave, Cambridge, MA 02139, USA.\\
\textsuperscript{\textdagger} {NASA Hubble Fellow.}
}

\begin{document}
\maketitle
\thispagestyle{empty}

\begingroup\footnotesize
\vspace{.3cm}
\noindent\textbf{Note:} This is the author’s version of the article. \\
The definitive version has been published in \textit{Nature} and can be accessed at \href{https://www.nature.com/articles/s41586-025-08760-2}{https://www.nature.com/articles/s41586-025-08760-2}.
\endgroup
%%%%%%%%%%%%%%%%%%%%%%%%%%%%%%%%%%%%%%%%%%%%%%%%%%%%%%%%%%%%%%%%%%%%%
% Abstract
%%%%%%%%%%%%%%%%%%%%%%%%%%%%%%%%%%%%%%%%%%%%%%%%%%%%%%%%%%%%%%%%%%%%%

\section{}
\small
\justifying

\noindent\textbf{
Acoustic oscillations in stars are sensitive to stellar interiors \cite{2021RvMP...93a5001A}. Frequency differences between overtone modes --large separations-- probe stellar density \cite{2022ApJ...927..167L}, while differences between low-degree modes --small separations-- probe the sound speed gradient in the energy-generating core of main sequence Sun-like stars \cite{2003A&A...411..215R}, and hence their ages. At later phases of stellar evolution, characterised by inert cores, small separations are believed to lose much of their power to probe deep interiors and simply become proportional to large separations \cite{2011ApJ...742L...3W,2017ApJ...835..172L}. 
Here, we present clear evidence of a rapidly evolving convective zone as stars evolve from the subgiant phase into red giants. By measuring acoustic oscillations in 27 stars from the open cluster M67, we observe deviations of proportionality between small and large separations, which are caused by the influence of the bottom of the convective envelope.
These deviations become apparent as the convective envelope penetrates deep into the star during subgiant and red giant evolution, eventually entering an ultra-deep regime that leads to the red giant branch luminosity bump.
The tight sequence of cluster stars, free of large spreads in ages and fundamental properties, is essential for revealing the connection between the observed small separations and the chemical discontinuities occurring at the bottom of the convective envelope.
We use this sequence to show that combining large and small separations can improve estimations of the masses and ages of field stars well after the main sequence.
}

%%%%%%%%%%%%%%%%%%%%%%%%%%%%%%%%%%%%%%%%%%%%%%%%%%%%%%%%%%%%%%%%%%%%%
\section{}
%%%%%%%%%%%%%%%%%%%%%%%%%%%%%%%%%%%%%%%%%%%%%%%%%%%%%%%%%%%%%%%%%%%%%
\subsection{Background}
\setstretch{1.3}
The oscillation spectra of Sun-like stars and their evolved counterparts, subgiants and red giants, originate from resonating acoustic pressure $(p)$ waves excited by surface convection (Figure~\ref{fig:F1}a).
 $p$-modes of spherical degree $\ell=0$ travel radially through the star and are reflected towards the core at the stellar surface. Non-radial $p$-modes (degree $\ell\geq1$) are refracted back to the surface and, therefore, are confined between an inner turning point and the surface. The radial coordinate of the inner turning point is a function of the spherical degree of the mode and the temperature gradient in the core.

Small separations refer to the frequency differences between modes of degrees $\ell$ and $\ell+2$, of consecutive order $n$. Due to the low visibility of modes of degree $\ell \ge3$ \cite{1995A&A...293...87K}, we focus on \smalltwo, the small separation between modes of degree $\ell=0$ and $\ell=2$. This separation is typically determined from models as a weighted average\cite{2011ApJ...742L...3W} of individual $\ell=0,2$ pairs, with weights determined by the frequency distance between the mode of degree $\ell=0$ and order $n$, and the frequency of maximum oscillation power, \numax.
 Asymptotic analysis\cite{1980ApJS...43..469T, 1990ApJ...358..313T, 1994MNRAS.267..297R} yields the approximate expression \smalltwo\ $\simeq -\frac{3}{\nu T}  \int_{0}^{R} \frac{dc_{s}}{dr} \frac{dr}{r}$ for a given frequency $\nu$, where $T$ is the acoustic radius, $R$ is the radius, $c_{s}$ is the sound speed, and $r$ is the radial coordinate. In main-sequence stars, this reduces to \smalltwo\ $\propto \sqrt{1/\mu}$ where 
 $\mu$ is the mean molecular weight\cite{2009IAUS..258..431C}, hence, %then 
 \smalltwo\ rapidly changes as the star burns hydrogen into helium. Therefore, in main sequence stars, small separations are a good indicator of evolutionary state\cite{1988IAUS..123..295C}, and hence age. Notably, the accuracy of this asymptotic approximation rapidly deteriorates as %the order $n$ 
 \numax\ decreases and the star becomes more centrally condensed during its evolution\cite{1990ApJ...358..313T}. 
Because of the lack of a suitable analytical expression for \smalltwo\ in subgiants and red giants, the relationship between small separations and the interior structure of stars is not fully understood after the main sequence.
A model-based interpretation has also been elusive 
due to the mixed nature of non-radial modes in subgiants and red giants, which result from the coupling between $p$-waves and $g$-waves (gravity waves trapped in the core)\cite{1977A&A....58...41A, 2013ApJ...767..158B}. This coupling produces irregular mode patterns and scatter in small separations\cite{2011ApJ...742L...3W}, making it difficult to obtain useful model predictions of small separations. Although the coupling weakens for late red-giant-branch stars\cite{2017A&A...600A...1M}, \smalltwo\ becomes nearly proportional to the large frequency separations, $\Delta\nu$, limiting the information it provides about the star\cite{2010ApJ...721L.182M, 2010ApJ...723.1607H, 2012ApJ...757..190C}.

\begin{center}
    \includegraphics[width=0.6\columnwidth]{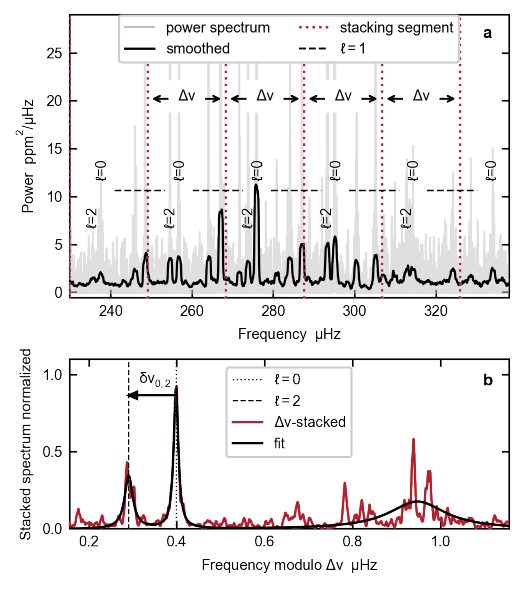}
\end{center}
\captionsetup{hypcap=false}\captionof{figure}{ \textbf{Oscillations in the red giant EPIC 211409560 in the open cluster M67}. a) In grey: region of the oscillation power spectrum centred around the frequency of maximum oscillation power. A slightly smoothed version of the spectrum is presented in black for clarity. The $\ell=0$ and $\ell=2$ modes are annotated, and the approximate ranges occupied by $\ell=1$ modes are indicated with horizontal dashed lines. Vertical dotted red lines indicate the segments used to stack the spectrum (Frequency modulo \dnu\ \muhz, in b). b) Stacked spectrum (red) and in black, the sum of three Lorentzian functions fitted to the stacked spectrum. The dotted and segmented black vertical lines mark the centres of the fitted Lorentzian profiles to the $\ell=0$ and $\ell=2$ modes, respectively. We measure the small frequency separation \smalltwo\ as the distance between these centres, as indicated by the arrow}
\label{fig:F1}
\vspace{0.25cm}

\subsection{Small frequency separations follow the structure of nuclear burning zones}

In Figure~\ref{fig:F2}a we show known critical stellar evolution points in the Hertzsprung-Russell diagram  (A, B, C, D, and G), which are associated with structural changes in the hydrogen burning regions seen in the Kippenhahn diagram (Figure~\ref{fig:F2}b). By using modelled pure $p$-modes\cite{2020ApJ...898..127O} isolated from the inner $g$-mode cavity we  produce evolution sequences of small versus large frequency separations, in so-called C-D diagrams\cite{1988IAUS..123..295C}, showing no mixed-mode induced scatter that would otherwise strongly distort the sequences\cite{2011ApJ...742L...3W}. The smooth C-D diagrams (Figure~\ref{fig:F2}c) now also show the inprint of all these critical points. Importantly, we see a new morphological feature in the CD diagram, which we call the plateau, bracketed by E and F. This feature, which has no counterpart in the other diagrams, appears during thin-shell burning as a temporary stalling in small frequency separations, while \dnu\ continues to decrease.

\begin{center}
\includegraphics[width=0.8\columnwidth]{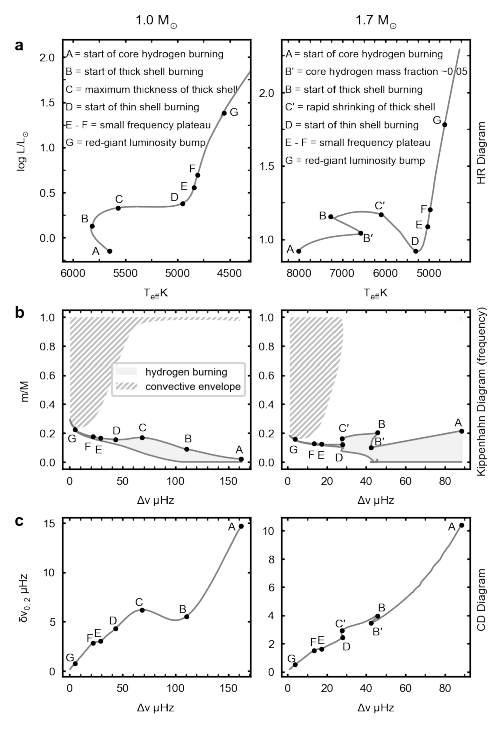}
\end{center}
\captionof{figure}{\textbf{Stellar tracks of solar metallicity}. (a) Hertzsprung\--Russell, (b) frequency-Kippenhahn, and (c) C-D diagrams of a 1.0 M$_\odot$ star (main sequence radiative core, left) and a 1.7 M$_\odot$ star (main sequence convective core, right). The black circle indicating point G, corresponding to the red giant luminosity bump, is covering the short lived fluctuation in all diagrams where the curves temporarily revert directions. In (b), regions where nuclear burning produces more than $10\, \mathrm{erg}\, g^{-1}\, s^{-1}$ are shown in light grey, and envelope convective regions are hatched.}
\label{fig:F2}
\vspace{0.25cm}

\subsection{The M67 plateau feature}
The near-solar-metallicity open cluster M67 (NGC 2682) presents a unique opportunity to investigate the nature of the plateau feature. This cluster has a rich subgiant and red giant population, which has been the target of attempted seismic studies for decades\cite{1993AJ....106.2441G}. Recent work includes a study of its giants\cite{2016ApJ...832..133S} using Kepler/K2 data\cite{Howell_2014}, which we also use in this study.

We analysed spectra from 27 shell-hydrogen-burning stars (Table~\ref{table:T1}) and determined their \smalltwo\ using a method that mitigates the influence of mixed modes (see \nameref{sec:Methods}), illustrated in Figure~\ref{fig:F1}b.
Figure~\ref{fig:F3}a shows the M67 C-D diagram, where the evolutionary state goes from subgiants (right) to red giants (left) as indicated by the black arrow. The models used to produce this pure $p$-mode C-D diagram correspond to a 3.95 Gyr theoretical isochrone specifically designed to provide the closest fit to M67 photometry to date\cite{2024MNRAS.532.2860R}.
The post-main-sequence section illustrated in Figure~\ref{fig:F3}a represents the evolved segment of the complete isochrone shown in Figure~\ref{fig:F3}b, and corresponds to stellar models in the mass range 1.30-1.37 M$_\odot$. 
In the observations, shown in black circles, we clearly detect the plateau where \smalltwo\ remains almost constant in the well-populated evolutionary locus of stars at $\Delta\nu$ between $\sim 17-22$ $\mu$Hz, indicated with a grey box in Figure~\ref{fig:F3}a. 
This feature, which is closely reproduced by the models, is evident in the data, and it probably remained undiscovered until now only due to the lack of a uniform sample of stars with similar fundamental properties (and hence, no intrinsic star-to-star scatter) needed to reveal it.

As indicated in the HR diagram in Figure~\ref{fig:F3}c with a grey box, 
the plateau occurs in stars as they ascend the red giant branch. In these stars, the electron-degenerate core contracts as it grows in mass, fed by the ashes of the hydrogen-burning shell. As per the mirror principle, the envelope expands and cools down, with the convective region deepening due to the increasing photospheric opacity of the cooling envelope\cite{2005essp.book.....S, 2017A&ARv..25....1H}.

\begin{figure*}[ht]
\centering
\includegraphics[width=0.5\textwidth]{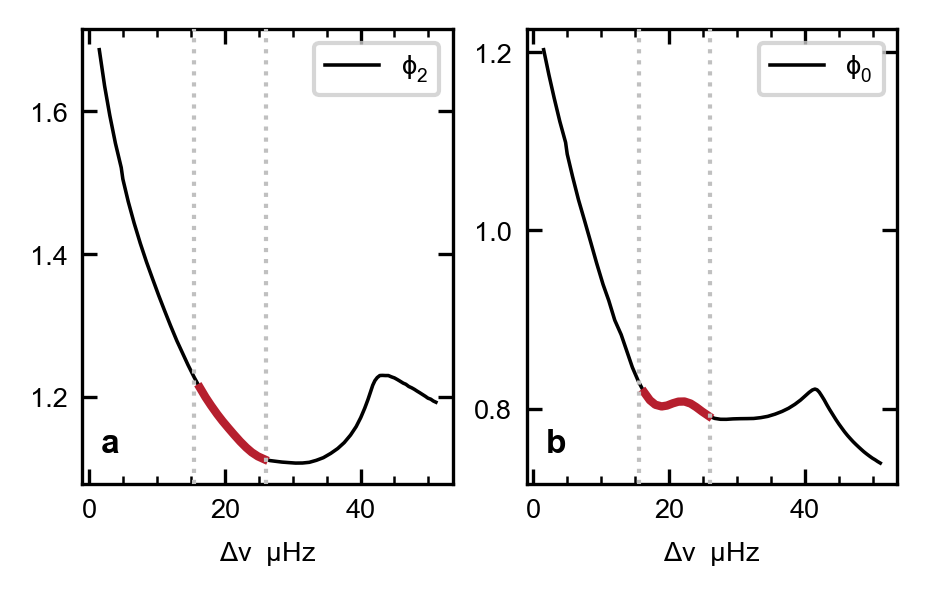}
\captionof{figure}{\textbf{Inner phase shifts in isochrone models.} Inner phase shifts $\phi_0$ and $\phi_2$ as averaged over modes near $\nu_\text{max}$ from models along the M67 isochrone \cite{2024MNRAS.532.2860R}. The dotted lines indicate the \dnu\ boundaries of Figure~\ref{fig:F3}e. Although no change in slope was observed in $\phi_2$ at the frequencies of the feature (a), a change in slope was found in the evolution of $\phi_0$ (b).}
\label{fig:S1}
\end{figure*}

We investigated the relative contributions of the $\ell=0$ and $\ell=2$ modes to the plateau 
in \smalltwo\ using the concept of internal phase shifts\cite{2003A&A...411..215R} $\phi_{\ell}$%\footnote{The notation $\delta_\ell$ is traditionally used for the inner phase function, but we use $\phi_\ell$ here to avoid confusions with \smalltwo.}
, given that ${\delta\nu_{02}} \sim {\frac{\Delta\nu}{\pi}}\left(\phi_2(\nu) - \phi_0(\nu)\right)$\cite{2005A&A...434..665R,2013A&A...560A...2R}. This is illustrated in Extended Data Figure~\ref{fig:S1}, where
the frequency range of the plateau in our observational C-D diagram is highlighted by the red sections. Within this range, the evolution of the quadrupole-mode inner phase $\phi_2$ can be seen to progress smoothly, while it is the evolution of the radial-mode inner phase $\phi_0$ that exhibits a local minimum, thereby producing the observed plateau. Thus, we conclude that the plateau in the observed C-D diagram probes stellar structural features lying near the star's centre, where only radial ($\ell=0$) $p$-modes reach, beyond the inner turning point of $\ell=2$ modes.

\subsection{The lower boundary of the convective envelope}

\begin{figure*}[ht]
\centering
\includegraphics[width=0.5\textwidth]{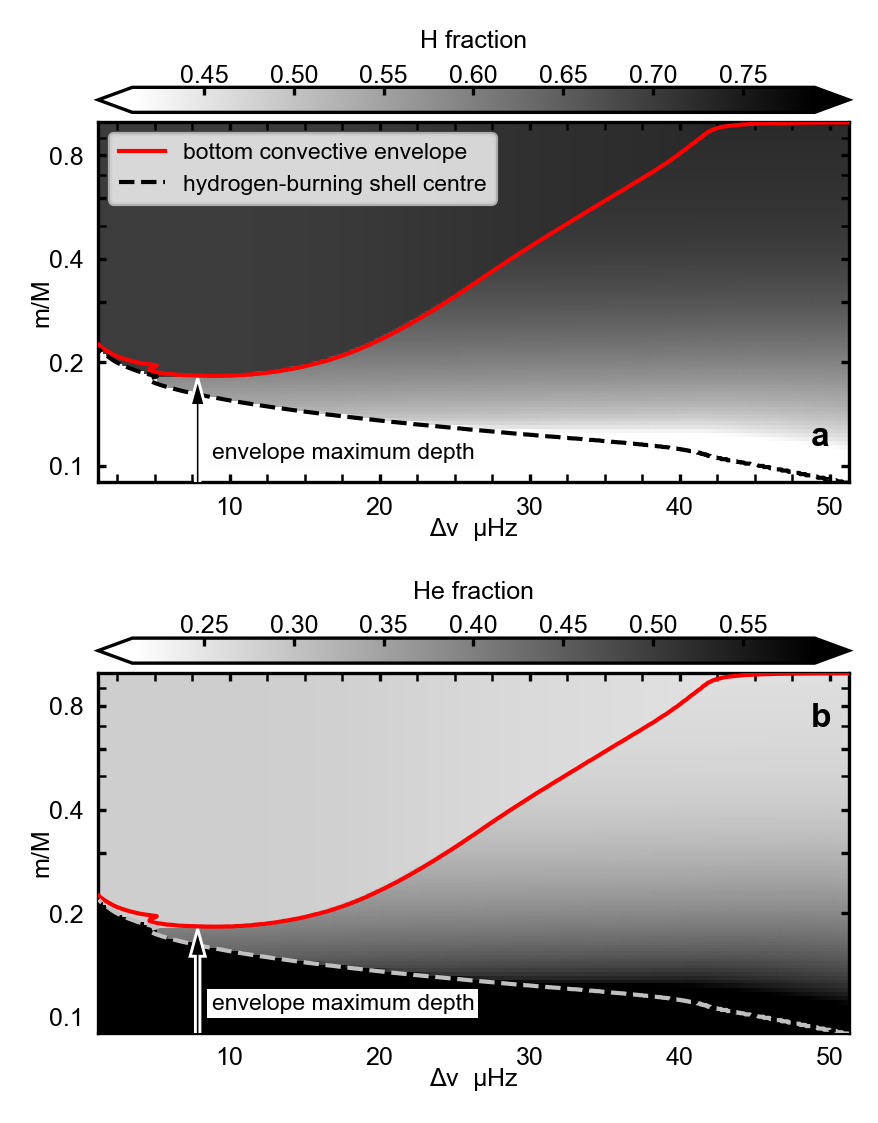}
\captionof{figure}{\textbf{Hydrogen and helium fractions and the bottom of the convective envelope.} Figures are equivalent to Figure~\ref{fig:F3}d, but with a grey scale showing the hydrogen and helium fractions. The bottom of the convective envelope (in red), traces the chemical discontinuity that leads to large gradients in molecular weight and temperature, as described in the main text.}
\label{fig:S2}
\end{figure*}

We find that the observed plateau can be traced to the evolution of the lower
boundary of the convective envelope. As the envelope expands and cools down, this lower boundary extends ever deeper into the stellar interior (Figure~\ref{fig:F2}b, Figure~\ref{fig:F3}d) as more efficient energy transport mechanisms are required deeper in the star.
Large density and sound-speed gradients are known to exist at such boundaries due to differing chemical compositions on either side, as shown in Extended Data Figure~\ref{fig:S2}. These gradients produce `acoustic glitches'\cite{2015ApJ...805..127C, 2017asda.book.....B, 2021RvMP...93a5001A, 2023ApJ...950...19L}, imparting frequency differences $\delta\nu_\text{glitch}$ compared to the mode frequencies of a smooth stellar structure with weaker gradients.
By writing the difference in density between the actual structure and such a smooth model as $\delta \rho$, the acoustic glitch signal may be described through expressions of the form $\delta\nu_{\text{glitch}, i} \sim \sum_q \int K_{q,i}\delta q(m)\ \mathrm d m$, where $m(r) = \int_0^r 4\pi r^2 \rho(r') \mathrm d r'$ is the mass coordinate, and $K_{q, i}$ is a sensitivity kernel associated with the quantity $q$ for the $i$\textsuperscript{th} mode. The average effect of $q$ on the radial-mode frequencies may then be examined by inspecting the averaged kernel $\left<K_q\right>$ over radial modes near $\nu_\text{max}$ (\nameref{sec:Methods}). 
For illustration, we take the amplitude of the density kernel $\left<K_{\rho, c_s^2}\right>$ (shown by the background colouring of Figure~\ref{fig:F3}d) along the position of the mixing boundary (solid black line) in mass coordinates, and we colour-code the seismic isochrone accordingly. In Figure~\ref{fig:F3}a, the C-D diagram appears to be modulated by the amplitude of the kernel at the mixing boundary. Furthermore, the plateau in the M67 C-D diagram occurs when the mixing boundary sweeps over one of the extreme points of this sensitivity kernel denoted by the darkest blue in Figure~\ref{fig:F3}. 

\begin{center}
\includegraphics[width=0.6\columnwidth]{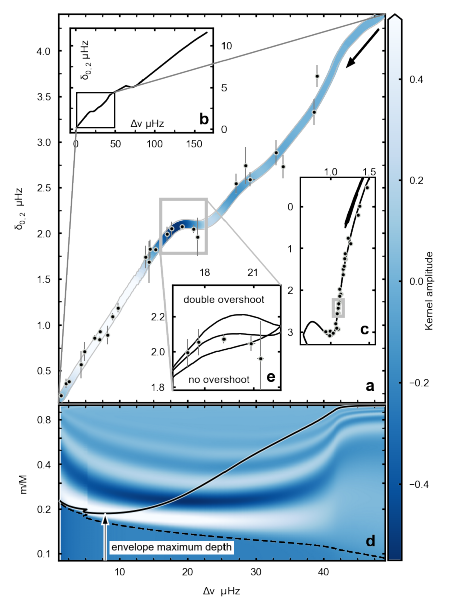}
\end{center}
\captionof{figure}{\textbf{Asteroseismic C-D diagram for subgiants and giants in the open cluster M67.} Panel a (which shares a horizontal axis with panel d) shows observed values of the large and small frequency separations for the M67 sample. In most cases the \dnu\ error bars are smaller than the symbols, indicating that \dnu\ uncertainties are negligible in this context. The \smalltwo\ errorbars were obtained as detailed in \nameref{sec:Methods}. The coloured curve represents a theoretical isochrone, colour-coded according to the amplitude of the averaged radial mode density kernel at the bottom of the convection zone, as shown in panel d. b) The theoretical C-D diagram corresponding to our sample of subgiants and red giants (within the grey box) in the context of the full C-D diagram, including the main sequence. c) The M67 sample and isochrone plotted in the HR diagram, with a grey box indicating the region corresponding to the grey box in panel a. d) \dnu\ and mass coordinates of the isochrone from panel a. The solid line indicates the bottom of the envelope convection zone, which reaches greater stellar depths as stars evolve from right to left; the segmented line marks the center of the burning shell, and the arrow points to the maximum depth of the envelope. Mass coordinates are colour-coded according to kernel amplitude. e) Close-up of the plateau feature of the M67 isochrone (central curve) compared to the same isochrone with double the envelope overshoot (upper curve) and with no envelope overshoot (lower curve)}
\label{fig:F3}
\vspace{0.25cm}

To verify the connection between \smalltwo\ and the bottom of the convection zone, we examine how the former changes when we vary the latter in our stellar modelling by altering the amount of convective boundary mixing in our computational treatment of stellar structure and evolution. In practice, we parameterise this convective boundary mixing as "convective overshooting", where convective motions extend beyond the nominal convective boundaries due to the inherent momentum of convective plumes\cite{1997ASSL..225...23R}, extending mixing regions, and thus relocating them relative to the regions of convective stability. 
 In Figure~3e we show two variants of the main isochrone:
one based on models with no envelope overshoot, and the other with twice the extent of overshoot compared to the solar calibrated overshoot factor\cite{2016ApJ...823..102C} of the adopted main isochrone. 
When we use models with more overshooting, the mixing boundary extends deeper into the radiative region compared to models with less or no overshooting. This means that the boundary will reach the critical kernel region earlier, and hence, we should see that with more overshooting, the deviation from proportionality occurs earlier in the evolution. Conversely, with no envelope overshoot the boundary takes longer to reach the same stellar depth, and we should see that the deviation from proportionality occurs later. Figure~3e confirms our predictions and further shows that with more overshooting, the plateau from M67 no longer presents a plateau, but a local maximum in \smalltwo\ that peaks at $\sim 20.5 \mu$Hz. With no overshooting, the feature is less prominent and has an inflection point at \smalltwo\ $\sim 18 \mu$Hz. A new theoretical expression for the small separations in red giants is necessary to fully explain the link between the prominence of the plateau feature and overshooting. However, by comparing with the data, we can say that the adopted solar-calibrated overshoot factor\cite{2016ApJ...823..102C} accurately predicts the correct convection zone depth in near-solar metallicity stars such as those in the M67 cluster. The location of the red giant branch luminosity bump \cite{2018ApJ...859..156K} and the evolutionary behaviour of the $\ell$ = 1 mixed modes \cite{2022ApJ...931..116L} near the luminosity bump are also dependent on the amount of overshooting at the bottom of the convection zone. Combined with these other indicators of convective envelope depth, the amount of envelope overshooting can now be analysed at multiple locations along the red giant branch, since the plateau feature in \smalltwo\ can sample the amount of envelope overshoot substantially before the luminosity bump.

\subsection{Mass dependence of plateau frequencies}

Our models also show that small separations behave similarly in other low-mass stars. Therefore, the plateau feature first observed in M67 provides a new diagnostic tool for determining stellar properties of field stars. 

\begin{figure*}[ht]
\centering
\includegraphics[width=0.5\textwidth]{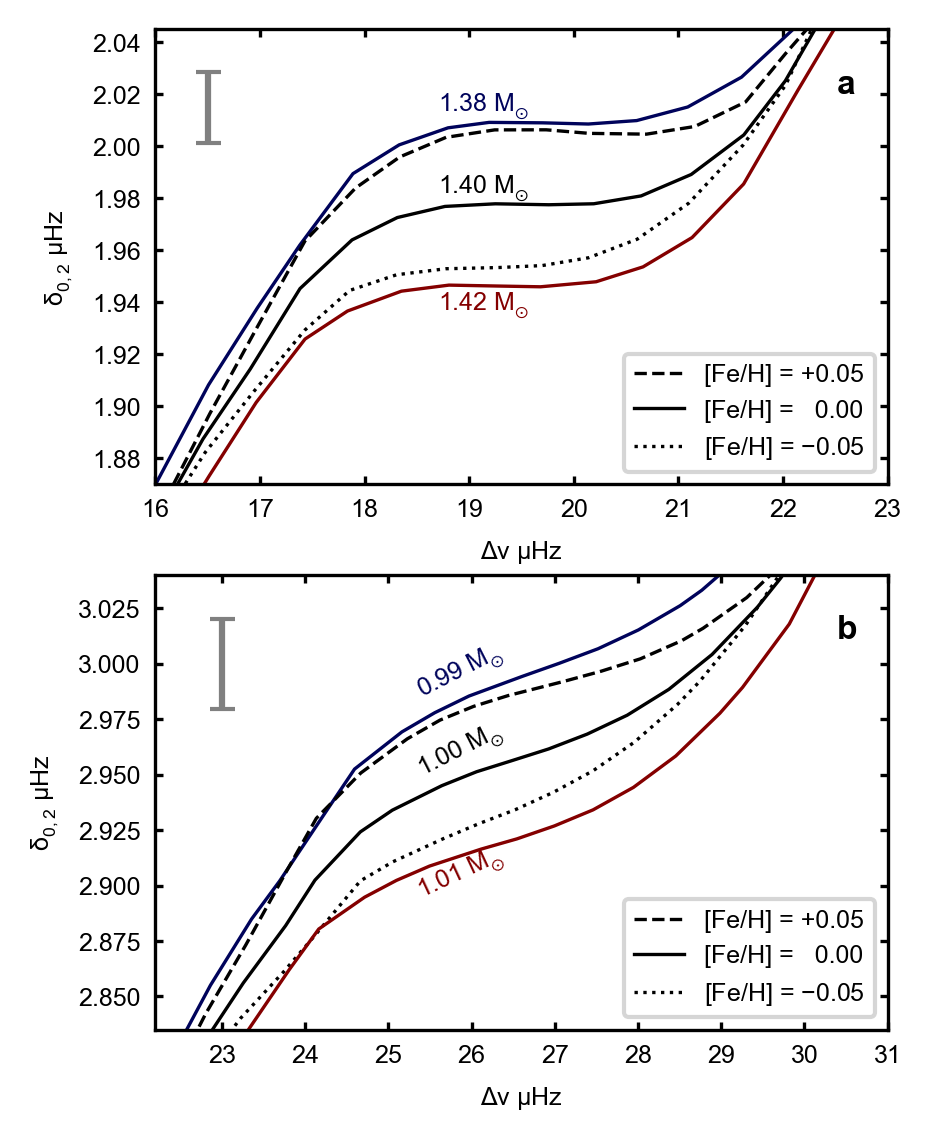}
\captionof{figure}{\textbf{Metallicity sensitivity of plateau features.} All solid lines correspond to [Fe/H]=0 tracks of masses as annotated. The dashed and dotted lines correspond to metallicity variations of the central track, of mass 1.4\msol, or 1.0\msol, as per the legend. The figures illustrate that translating a typical [Fe/H] uncertainty, estimated at 0.05 to 0.10 dex\cite{2016AJ....151..144G, 2020AJ....160..120J}, into the plateau indicates that for solar-metallicity tracks at 1.40 \msol, this uncertainty corresponds to less than 0.02 \msol. For solar-metallicity tracks at 1.00 \msol, the [Fe/H] uncertainty translates to less than a 0.01 \msol\ uncertainty. Typical \kepler\ error bars for small separations of 2 \muhz\ and 3 \muhz\ are shown next to the 1.4\msol\ and 1.0\msol\ tracks, respectively (see \nameref{sec:Methods}).}
\label{fig:S3}
\end{figure*}

\begin{figure*}[ht]
\centering
\includegraphics[width=\textwidth]{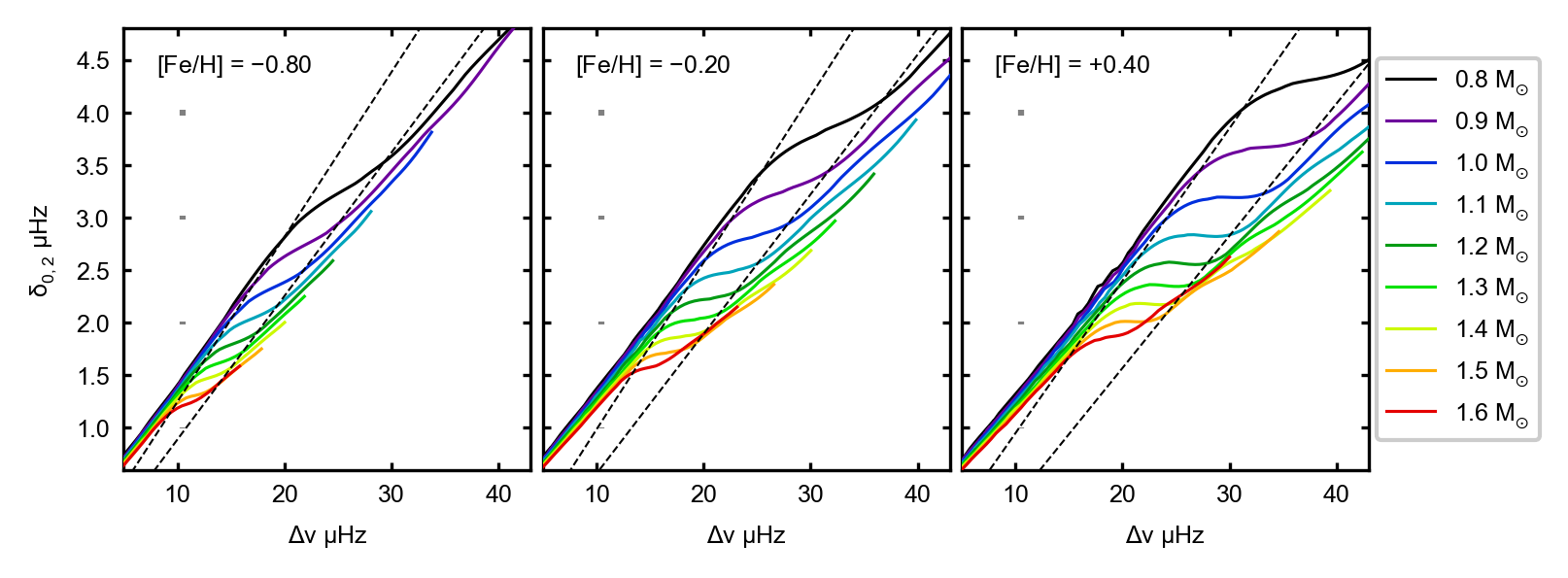}

\caption{\textbf{Plateau-like features across different metallicities.} The figures illustrate the metallicity dependence of plateau-like features across [Fe/H] from -0.8 to +0.4 dex, showing masses from 0.8\msol\ to 1.6\msol. Dotted lines approximately trace the region where we see the plateau for each metallicity. 
The figure illustrates the shift of the plateaus from the lower left to the upper right corner of the diagram, with the progression from the most metal-poor to the most metal-rich tracks. For the latter, flatter and wider plateaus emerge. 
Typical \kepler\ uncertainties (see \nameref{sec:Methods}) are indicated at \dnu=10 for \smalltwo=2, 3, and 4 \muhz.}
\label{fig:S4}
\end{figure*}

\begin{figure*}[ht]
\centering
\includegraphics[width=0.35\textwidth]{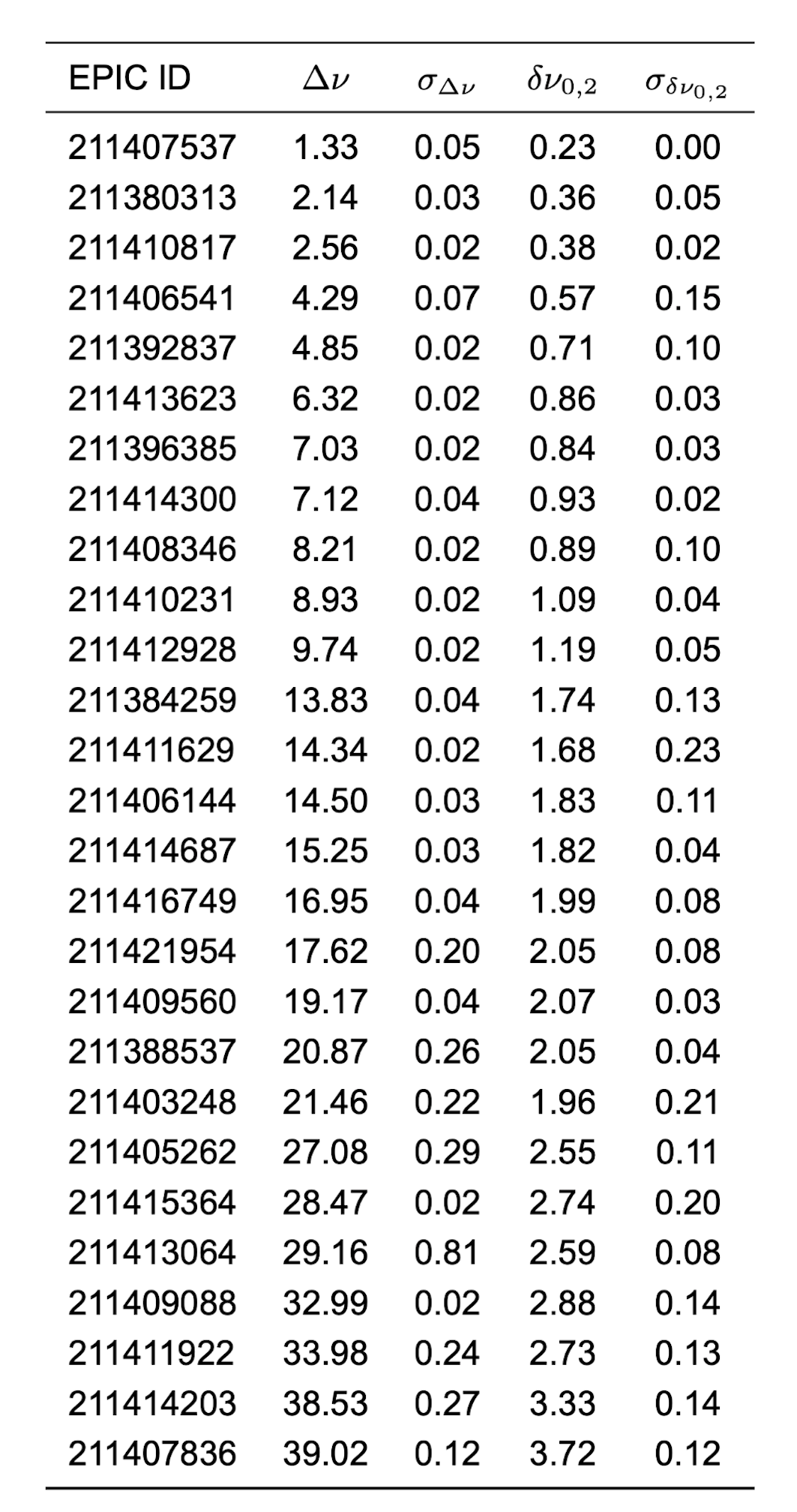}
\captionof{table}{\textbf{Seismic parameters of M67 stars.} Stars from M67 with measured small frequency separation $\delta\nu_{0,2}$. 
The uncertainty $\sigma_{\Delta\nu}$ is taken from results by the pysyd pipeline. \dnu, \smalltwo, and $\sigma_{\delta\nu_{0,2}}$ 
are calculated as described in \nameref{sec:Methods}. All values are given in $\mu$Hz.}
\label{table:T1}
\end{figure*}

Figure~\ref{fig:F4} shows a C-D diagram derived from stellar tracks in the range $0.8-1.6\,\mathrm{M}_{\odot}$, with solar metallicity, starting at the 
%Terminal-Age Main Sequence 
beginning of core hydrogen burning to just before core helium burning, or until a stellar age of 12 billion years, whichever is first. As before, we achieved this level of detail by calculating modelled frequencies using only pure $p$-modes\cite{2020ApJ...898..127O}. A plateau feature is well-defined in all tracks shown, and appears at values of \smalltwo\ that are specific for each track. 
Therefore, by placing observations of \smalltwo\ and \dnu\ on grids built from these models, this new diagnostic tool could be used to accurately estimate the masses of field red giants. % in the \dnu\ range of the \smalltwo\ plateaus,
This is illustrated by the inset in Figure~\ref{fig:F4}b, which shows typical \kepler\ \smalltwo\ uncertainties --and negligible \dnu\ uncertainties-- (see \nameref{sec:Methods}),  between the plateaus of the 1.4 and 1.5 \msol\ tracks. Additional details on the metallicity dependence of the plateau frequencies are provided in Extended Data Figure~\ref{fig:S3} and Figure~\ref{fig:S4}. 
Outside the \dnu\ range of the plateau, towards more evolved giants, the diagnostic power of the post-main sequence C-D diagram is diminished because the tracks converge. 

\begin{center}
    \includegraphics[width=0.6\columnwidth]{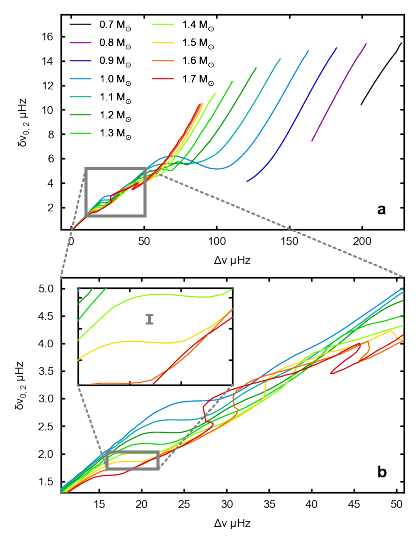}
\end{center}
\captionof{figure}{\textbf{The C-D diagram of a sequence of solar metallicity stellar tracks.} a) Stellar tracks between 0.7 and 1.7 M$_{\odot}$ starting at the beginning of core hydrogen burning and ending just before the helium flash or at a stellar age of 12 billion years. The grey box in a) indicates the section shown in detail in b), where the plateau features are clearly discernible at all masses shown. The inset in b) shows a typical \smalltwo\ uncertainty for a star between 1.4 and 1.5 \msol\ observed by \kepler, while typical \kepler\ \dnu\ uncertainties are negligible in this context.}
\label{fig:F4}
\vspace{0.25cm}

\subsection{M67 reveals structural changes in deep stellar interiors}

As the first single stellar population with clear measurements of small frequency separations across the subgiant and red giant branches, M67 reveals that the depths reached by convective envelopes lead to measurable effects in the small-separation frequencies, which reveals itself as a plateau in the C-D diagram. At the end of the plateau, the convective envelope enters its ultra-deep regime, beginning when roughly 80\% of the star's mass is undergoing convection. This fraction continues to increase until the convective envelope reaches its maximum depth. The envelope then retreats, leaving behind a chemical discontinuity imprint. Eventually, this discontinuity is erased as the shell burns through it in what is known as the red giant luminosity bump --until now thought to be the only observational evidence of the depth reached by the convective envelope.
The well-defined distribution of plateau frequencies according to mass, combined with the strong mass-age relation for giants\cite{2012ASSP...26...11M} and the relative ease of measuring small separations, also make this feature interesting for age-dating red giants in the field and, hence, for mapping the chronology of the Milky Way merger events\cite{2021A&A...645A..85M}.

%%%%%%%%%%%%%%%%%%%%%%%%%%%%%%%%%%%%%%%%%%%%%%%%%%%%%%%%%%%%%%%%%%%%%
\section{Methods}
\label{sec:Methods}

%%%%%%%%%%%%%%%%%%%%%%%%%%%%%%%%%%%%%%%%%%%%%%%%%%%%%%%%%%%%%%%%%%%%%

We describe here the seismic analysis of 27 M67 stars observed by the K2 mission, the stellar models we developed for the cluster, and the theory behind phase shifts and kernels.

\subsection{Oscillation spectra and seismic characterisation}

\subsubsection{M67 K2 data}
We downloaded all the available K2 light curves of M67 cluster members\cite{2024MNRAS.532.2860R} from the Mikulski Archive for Space Telescopes (MAST, https://archive.stsci.edu), corrected them\cite{2016ApJ...832..133S} and calculated their power density spectrum\cite{1976Ap&SS..39..447L}.
We obtained the initial values of \numax\ and \dnu\ and background properties using pysyd \cite{2021ascl.soft11017C}, which is a python implementation of the SYD\cite{2009CoAst.160...74H} pipeline. First, we calculated the \dnu-stacked spectrum 
by averaging the 4 \dnu-wide segments closest to \numax\ of the background-subtracted power spectrum. We then fitted the sum of three Lorentzian functions to this stacked spectrum as in Figure \ref{fig:F1}b. This spectrum stacking method boosts the signal-to-noise ratio of $\ell=0$ and $\ell=2$ modes while also reducing the impact of mixed modes in the final \smalltwo\ measurements. 
To quantify the uncertainty in \smalltwo, we calculated the fractional differences between the widths and signal-to-noise ratios of the Lorentzians fitted for $\ell=0$ and $\ell=2$ modes. After summing the squares of these differences and taking the square root of the sum, we incorporated the resulting value with the propagated uncertainties from the individual fits.
We found that this combined uncertainty is sensitive to missing modes and unusually high signal modes, which are the primary contributors to inaccuracies in \smalltwo\ in relatively short lightcurve data, such as K2 data.
To refine pipeline \dnu\ determinations, we sought the \dnu\ value that maximised the height and minimised the width of the Lorentzian fit to the $\ell=0$ peak.
Finally, we rejected stars whose \dnu-stacked $\ell=0$ or $\ell=2$ peaks had a signal-to-noise ratio lower than 3, resulting in a sample of 27 stars. Values are presented in Table~\ref{table:T1}.

\subsubsection{\kepler\ large and small separation uncertainties}
We calculate an average fractional uncertainty of 0.05\% in \dnu\ and 0.7\% in \smalltwo\ based on measurements of 188 \kepler\ red giants within the range 15<\dnu<20 \muhz\cite{2019arXiv190609428K}.

\subsection{Stellar models}

\subsubsection{Tracks and profiles}
To generate the isochrone models with no (or double) envelope overshoot from Figure~\ref{fig:F3}e, we adapted models from the M67 isochrone\cite{2024MNRAS.532.2860R}. We note that this isochrone used a mass-dependent core overshoot, and a fixed solar-calibrated envelope overshoot, both using the exponential overshoot scheme, and H and He content as documented\cite{2024MNRAS.532.2860R}. 
We adapted the same models to generate the $0.8-1.6 \mathrm{M}_{\odot}$ tracks from Figure~\ref{fig:F4}, and Extended Data Figure~\ref{fig:S3} and Figure~\ref{fig:S4}, except that the tracks shown in these Figures use the solar H and He fractions from Asplund 2009 as reference.

\subsubsection{Radial $\ell=0$ and non-radial $\ell=2$ \textit{p}-mode frequencies} 
We calculated adiabatic frequencies from structure profiles using the oscillations code GYRE\cite{2013MNRAS.435.3406T} version 6.0.1. To obtain the smooth sequence of the C-D diagram, we required mode frequencies free of any $g$-mode quality. Because only $\ell=0$ modes are intrinsically independent of any influence from $g$-modes, we applied a formalism based on semi-analytic expressions for the isolation of modes\cite{2020ApJ...898..127O} to compute $\ell=2$ pure $p$-modes: the pulsation equations are decomposed into a pure $p$-mode wave operator and a remainder term from the radiative interior. The eigenvalues of the former are solved, and pure $p$-mode frequencies are recovered by applying perturbation theory to the latter.

\subsubsection{Modelled seismic data}

Surface corrections are required\cite{2014A&A...568A.123B, 2017ApJ...835..173S} to help minimise the impact of poor modelling of the outer layers in 1D stellar evolution codes before one can compare models to observed frequencies.
For all models, we used a smooth form of surface correction that follows the corrections to radial modes of individual stars in our sample\cite{2014A&A...568A.123B}. It is a reasonable approach to apply the same surface offset for both $\ell=0$ and $\ell=2$ modes, given that we exclusively work with pure $p$-modes. 
To obtain \dnu\ for all the selected models, we weighted the $\ell=0$ frequencies by a Gaussian window of width 0.25\numax, centred on \numax, and performed a least-squares fit to the frequencies as a function of mode order $n$, where the slope of this fit is \dnu\cite{2011ApJ...743..161W}.
Small separations \smalltwo\ are calculated weighting $\nu_{0,n} - \nu_{2,n-1}$ by the same Gaussian window as before, now performing a least-squares fit to $\nu_{2,n-1}-\nu_{\mathrm{max}}$, and extracting the intercept of the fit\cite{2011ApJ...743..161W}.

\subsubsection{Inner phase shifts}
We calculate the inner phase shift\cite{Roxburgh2010} of a particular mode, $\phi_{\ell}$, as a function of the acoustic radius, $t = \int_0^r dr / c_s(r)$, by evaluating
\begin{equation}
    \phi_{\ell}(t) = \tan^{-1} \bigg( \frac{\omega \psi}{d\psi / d t} \bigg) - \omega t + \frac{\pi}{2}\ell
\end{equation}
at location $t = 0.5\mathrm{T}$, for both the radial ($\ell = 0$) and quadrupole ($\ell = 2$) modes, where T is the acoustic radius at the model's surface, $\omega$ is the angular mode frequency and $\psi = r p' / \sqrt{c_s \rho}$, with $r$ being the radius, $c_s$ the sound speed, $\rho$ the density, and $p'$ is the Eulerian pressure perturbation of the mode.
For each stellar model and degree, we evaluate the inner phase shifts for all modes
\cite{2003A&A...411..215R}, then perform a weighted average over the frequencies using a Gaussian window\cite{2011ApJ...742L...3W} centred at $\nu_{\text{max}}$ with a full-width at half-maximum of\cite{Mosser2012} 
\begin{equation}
    \Gamma = 0.66\ \mu\mathrm{Hz} \times (\nu_\text{max}/\mu \mathrm{Hz})^{0.88}.\label{eq:mosser} 
\end{equation}

\subsubsection{Density kernel}

Sharply-localised structural features in the stellar interior perturb p-mode frequencies from the uniform spacing predicted by their asymptotic relation. For p-modes in particular, features in the density and the sound speed in particular yield such frequency perturbations through integrals against localisation kernels:
\begin{equation}
    {\frac{\delta\omega_i}{\omega_i}} \sim \int K_{\rho,c_s^2,i} {\frac{\delta\rho}{\rho}}\ \mathrm d r + \int K_{c_s^2,\rho,i} {\frac{\delta c_s^2} {c_s^2}}\ \mathrm d r,
\end{equation}
where $\delta\rho$ and $\delta c_s^2$ indicate departures in the density and sound-speed profiles of the star from, say, a smoothly stratified polytrope. We compute these as\cite{1991sia..book..519G}
\begin{equation}
\begin{aligned}
    K_{c_s^2,\rho}(r) &= {\rho c_s^2 \chi^2 r^2 \over 2 I \omega^2};\\
    K_{\rho,c_s^2}(r) &= {\rho r^2 \over 2 I \omega^2} \Bigg[c_s^2 \chi^2 - \omega^2\left(\xi_r^2 + \Lambda \xi_h^2\right) - 2 g \xi_r \chi \\ & - 4\pi G\int_r^R \xi_r\left(2\rho\chi + \xi_r {\mathrm d \rho \over \mathrm d r}\right)\ \mathrm d r'
    \\ & + 2 g \xi_r {\mathrm d \xi_r\over \mathrm dr} + 4\pi G \rho \xi_r^2 + 2\left(\xi {\mathrm d \Phi\over \mathrm d r} + \Lambda \xi_h {\Phi \over r}\right) \Bigg],\\ & 
\end{aligned}
\end{equation}
where $\xi_r$ and $\xi_h$ are the radial and horizontal components of the Lagrangian displacement $\vec{\xi}$ of the mode, $\chi = (\nabla \cdot \vec\xi) / Y_\ell^m$, $\Lambda = \ell(\ell + 1)$, $g=Gm/r^2$ is the local gravitational field, and $\Phi$ is the perturbation to the gravitational potential. These two kernels are offset from each other by a phase lag of $\pi \over 2$. The averaged kernel depicted in \ref{fig:F2} was then constructed by averaging $K_{\rho, c_s^2}$ over all radial orders near $\nu_\text{max}$ with weights given by a Gaussian envelope centred on $\nu_\text{max}$ with width given by \ref{eq:mosser}.

%%%%%%%%%%%%%%%%%%%%%%%%%%%%%%%%%%%%%%%%%%%%%%%%%%%%%%%%%%%%%%%%%%%%%
\section{Acknowledgments}
%%%%%%%%%%%%%%%%%%%%%%%%%%%%%%%%%%%%%%%%%%%%%%%%%%%%%%%%%%%%%%%%%%%%%

D.S. is supported by the Australian Research Council (DP190100666).
J.O. acknowledges support from NASA through the NASA Hubble Fellowship grant HST-HF2-51517.001, awarded by STScI. STScI is operated by the Association of Universities for Research in Astronomy, Incorporated, under NASA contract NAS5-26555.

%%%%%%%%%%%%%%%%%%%%%%%%%%%%%%%%%%%%%%%%%%%%%%%%%%%%%%%%%%%%%%%%%%%%%
\section{Author Contributions}
%%%%%%%%%%%%%%%%%%%%%%%%%%%%%%%%%%%%%%%%%%%%%%%%%%%%%%%%%%%%%%%%%%%%%
Author Contributions: C.R. calculated power spectra and measured frequency separations; J.O. and C.L. calculated inner phase shifts and kernel functions; 
C.R. and C.L. calculated theoretical models; D.S. and C.R. performed the analysis; C.R., D.S., J.O. and C.L. generated figures and participated in writing the manuscript; All authors discussed the results and commented on the manuscript.

\section{Author Competing Interests}
The authors declare no competing interests.

\section{Data availability}
K2 light curves are available from \href{https://archive.stsci.edu/}{MAST}. Power spectra, isochrone and stellar tracks are available at
\href{https://zenodo.org/records/12617071}{Zenodo}.

\section{Additional Information}
Correspondence and requests for materials should be addressed to claudiarreyes@icloud.com. Reprints and permissions information is available at www.nature.com/reprints.

\clearpage

%%%%%%%%%%%%%%%%%%%%%%%%%%%%%%%%%%%%%%%%%%%%%%%%%%%%%%%%%%%%%%%%%%%%%
\beginsupplement %
\section{Extended Data}
\label{sec:supplementary}

\newpage
\end{document}